# Evidence for a universal length scale of dynamic charge inhomogeneity in cuprate superconductors

D. Mihailovic

Jozef Stefan Institute, Jamova 39, 1000 Ljubljana, Slovenia

Time-resolved optical experiments can give unique information on the characteristic length scales of dynamic charge inhomogeneity on femtosecond timescales. From data on the effective quasiparticle relaxation time $\tau_r$ in $La_{2-x}Sr_xCuO_4$ and $Nd_{2-x}Ce_xCuO_4$ we derive the temperature- and doping- dependence of the intrinsic phonon escape length $l_e$, which, under certain circumstances, can be shown to be a direct measure of charge inhomogeneity. Remarkably, a common feature of both $p$ and $n$-type cuprates - which has important consequences for superconductivity - is that as $T \rightarrow T_c$ from above, the escape length approaches the zero-temperature superconducting coherence length, $l_e \rightarrow \xi_s(0)$. In close vicinity of $T_c$, $l_e$ appears to follow the critical behaviour of the Ginsburg-Landau coherence length, $\xi_{GL}(T)$. In the normal state $l_e$ is found to be in excellent agreement with the mean free path $l_m$ obtained from the resistivity data. The data on $l_e$ also agree well with the data on structural coherence lengths $l_s$ obtained from neutron scattering experiments, implying the existence of complex intrinsic textures on different length scales which may have a profound effect on the functional properties of these materials.

In recent years, substantial experimental evidence has started to accumulate for the existence of inhomogeneity in cuprates and other functional oxides, suggesting that it plays a fundamental role in their physics. Unfortunately, *dynamic* inhomogeneity – which is particularly important - is not easily and unambiguously detectable[1] since many standard techniques for investigating electronic and lattice structure fail when it comes to detecting dynamically changing deviations from the average structure. The current evidence for dynamic structural inhomogeneity comes from XAFS[2] and neutron PDF[3] experiments, which have sufficiently short intrinsic experimental timescales to give information on dynamical local structure fluctuations, but unfortunately have a limited range of – at most - a few lattice constants[4]. Some other experiments, such as inelastic neutron scattering[5,6,7] and scanning tunnelling microscopy[8] have suggested a characteristic length scale of the inhomogeneity of the order of 1-2 nanometers, but no information on the dynamics can be deduced from these experiments.

In this report, we present new femtosecond-timescale information on charge inhomogeneity as a function of temperature and doping in selected cuprate superconductors. Using time-resolved optical experiments we extract the characteristic length scale $l_e$ of charge inhomogeneities from the effective quasiparticle lifetime $\tau_r$, and compare these results to data obtained from other established techniques.

We find remarkable agreement of the dynamical length scale measured by time-resolved spectroscopy with structural coherence lengths $l_s$ determined from inelastic diffuse neutron scattering as well as the normal state mean free path $l_m$ obtained from conductivity measurements over a wide range of temperatures.

To understand how we can derive a length scale from time-resolved measurements of the quasiparticle lifetimes, let us first consider the dynamics of the creation of bound states such as Cooper pairs (or charged stripes) from unbound particles. The formation of a bound state from unbound Fermions leads to the release of energy $2\Delta$ in the form of a phonon (there is no other excitation which can remove the energy from the electronic subsystem). However, as was pointed out by Rothwarf and Taylor (R-T) already in 1967[9], unless the phonon lifetime is zero, the emitted phonon can be efficiently reabsorbed again, breaking pairs, and leading to the formation of a relaxation bottleneck. The overall kinetics of pairing are typically described by the R-T rate equations for the quasiparticle density $n$, and the optical phonon density $N$ respectively as $\frac{dn}{dt} = I_0 + \beta N - Rn^2$ and $\frac{dN}{dt} = -\frac{1}{2}(\beta N - Rn^2) - \gamma(N - N_T)$, where $\beta$ is the pair-breaking rate by phonons with $\hbar\omega_{ph} > 2\Delta$, $R$ is the recombination rate, $N$ is the non-equilibrium density and $N_T$ is the density of optical phonons in thermal equilibrium. $\gamma$ is the rate at which optical phonons decay or escape by processes other than pair excitation. Under bottleneck conditions, $\beta N \approx Rn^2$, and the overall relaxation rate is dominated by the phonon escape term $\gamma(N-N_T)$. Thus, the experimentally observed relaxation time $\tau_r$ is determined by the escape of phonons from the recombination volume, rather than by the direct pair recombination rate $R$. In low-$T_c$ superconductors which have a relatively small superconducting energy gap $2\Delta < 10$ meV, acoustic phonons are emitted by this process, and experiments on Al and Pb have clearly shown that the observed recombination time $\tau_r$ depends on superconducting film thickness, as predicted by the R-T model[10].

In HTS superconductors, pre-formed pairs or other forms of bound states may form in the normal state ($T > T_c$), so R-T relaxation might be expected also above $T_c$, where it would apply to the formation of bipolarons or stripes from unbound particles. First, high frequency optical phonons (with $\hbar\omega > 2\Delta$) are emitted, which subsequently decay anharmonically to acoustic phonons[11]. However, these acoustic phonons can efficiently recombine again to excite optical phonons, which in turn break up pairs again until eventually the acoustic phonons diperse out of the recombination volume. By analogy with the R-T equations, the rate equations for the optical phonon decay to acoustic phonons and acoustic phonon escape can then be written as:

$$\frac{dN}{dt} = G_0 - \frac{1}{2}(\chi N - \delta M^2) \qquad (1)$$

$$\frac{dM}{dt} = \chi N - \delta M^2 - \kappa(M - M_T) \qquad (2)$$

Where $G_0$ is the optical phonon generation rate by QPs, $M$ is the density of acoustic phonons, and $\kappa(M - M_T)$ describes the acoustic phonon escape, $\chi$ is the rate of optical phonon decay to acoustical phonons, while $\delta$ is the rate of acoustic phonon re-absorption. Under bottleneck conditions (i.e. near-steady-state) $\chi N \approx \delta M^2$, so using $\frac{dM}{dt} = v_s \frac{dM}{dr}$, Eq. (2) we obtain an explicit equation for the phonon escape:

$$\frac{dM}{dr} = -l_e^{-1}(M - M_T) \qquad (3)$$

where $l_e = v_s/\kappa$ is the phonon escape length. The justification for the steady-state assumption can be tested through comparison with experiments. If it can be shown that $l_e$ is determined by geometrical constraints, such as film thickness, as was done in conventional superconductors, then the steady-state assumption can be fully justified. Importantly, if the material is intrinsically inhomogeneous, the relevant geometrical constraints on the recombination volume are determined by the inhomogeneity length scale, and the relaxation time may be dominated by the phonons' escape from one phase (charge-rich) into another (charge-poor), as shown schematically in the insert to Fig.1. Inhomogeneities on the nanometer scale give rise to phonon escape times on the picosecond timescale, which is directly measurable with current ultrafast laser spectroscopy.

Over the past few years, several experiments have been carried out on cuprates to measure the temperature dependence of the relaxation time $\tau_r$ [11,12,13,14,15,16,17]. A typical data set showing the photoinduced normalized reflectivity as a function of time is shown in Figure 1 for $La_{2-x}Sr_xCuO_4$, at different temperatures for $x$=0.1. (The experimental details are described in ref. 18). From the fitted relaxation time $\tau_r$, we can analyse the characteristic phonon escape lengths $l_e$ as a function of $T$. We concentrate on $La_{2-x}Sr_xCuO_4$ (LSCO) here, because the relevant data are most complete for this compound. Using measured values of $\tau_r$[18] and $v_s$[19], in Figure 2 we have plotted $l_e$ for x = 0.06 - 0.2, at different temperatures. For all superconducting samples ($x$ = 0.1-0.2), $l_e$ shows a gradual increase from approximately $l_e \approx 1$ nm at 300K to about $l_e \approx 2 – 3$ nm near $T_c$. Well below $T_c$, $l_e$ continues to increase rapidly, but with a different slope. (The insulating sample on the other hand (x=0.06) shows almost no temperature dependence of $l_e$ over this temperature range.)

Close to $T_c$, $l_e$ shows signs of critical behaviour, exhibiting a clear peak. This peak is more pronounced in some materials than others[11,12,13,18], and depends on experimental conditions (such as incident laser power), but it is generally present only in a narrow 2-3K region near $T_c$. At a 2$^{nd}$ order phase transition (such as the superconducting transition), the correlation (coherence) length is expected to diverge as $\xi_s \sim (T_c-T)^{-\nu}$ above and as $\xi_s \sim (T-T_c)^{-\nu}$ below $T_c$. Plotting the critical region near $T_c$ for $La_{1.9}Sr_{0.1}CuO_4$ in the insert to Fig. 3, we see that the critical exponent of $l_e$ is near $\nu=0.5$, the mean field theory value, and is consistent with the relation $l_e = v_s \tau_r \sim \Delta^{-1} \sim \xi_s^{-1/2}$ discussed previously[13]. The obvious implication from the data in Fig. 3a), that $l_e$ is related to $\xi_s(T)$ in the vicinity of $T_c$ is quite remarkable, and is far from trivial. Although the detailed theoretical justification for this statement is beyond the scope of this paper, we wish to remark here that the connection between $\xi_s(T)$ and $l_e(T)$ implies that the charge inhomogeneity governs the behaviour of the superconducting phase coherence and the transition to the superconducting state at $T_c$[20]. The corollary is that the presence of inhomogeneity on a scale shorter than the coherence length clearly prevents the formation of a phase coherent superconducting state.

We can check if our understanding of the recombination dynamics in terms of inhomogeneity-limited phonon escape is appropriate, by comparing $l_e$ with the length scales obtained from other experiments which can sense dynamic inhomogeneity. An obvious example is the mean free path $l_m$ of charge carriers in the normal state, which can be obtained from DC resistivity measurements. Expressing $l_m$ via the Drude formula in terms of the relevant experimentally measurable quantities such as the *ab*-plane superconducting penetration depth $\lambda_s$ and Fermi velocity $v_F$ (rather than effective mass and density), we obtain $l_m = \frac{\lambda_s^2 v_F}{\varepsilon_0 c^2} \sigma(T)$, where $\sigma(T)$ is the normal state conductivity. In Figure 3a) we show a comparison between $l_e$ and $l_m$ (normalised to the superconducting coherence length $\xi_s$) using measured values of $\lambda_s$[21], $v_F$[22] and in-plane $\sigma(T)$[23] for $x = 0.1$ and $x = 0.15$ as a function of $T/T_c$. We observe remarkable quantitative agreement between $l_e$ and $l_m$ over the entire temperature range ($T_c < T < 300K$) where the data sets overlap. The implication is that the charged carrier mean free path $l_m$ and the phonon escape length $l_e$, are both being determined by the same underlying texture, and empirically, $\frac{\sigma(T)}{\tau_r(T)} \approx \frac{\varepsilon_0 c^2 v_s}{\lambda_S^2 v_F}$.

To determine if the underlying length scale is connected with *structural* inhomogeneity, we compare $l_m$ and $l_e$ with the *structural* coherence length $l_s$ determined by neutron scattering experiments on LSCO. The latter is defined as the length scale over which a particular structure which gives rise to the diffraction peak is coherent[7]. Although we are not aware of any detailed studies of structural inhomogeneities over a large range of temperatures, Kimura et al[7] deduced a structural coherence length of $l_s = 1/\Delta k \sim 1.7$ nm in slightly underdoped $La_{1.88}Sr_{0.12}CuO_4$ around 315 K from the linewidth of the diffuse central peak ($\Delta k=0.06$ Å$^{-1}$) in neutron scattering. Well below $T_c$ (at $T=13.5$ K), this structural coherence length is apparently extended to $l_s > 10$ nm[7]. Inelastic neutron scattering experiments on phonons in LSCO also show an anomaly with a characteristic wavevector $k_0 \sim \pi/2a$ (where $a$ is the lattice constant), corresponding to a length scale of the underlying textures of $l \approx 2$ nm [5]. Comparing these values with $l_e$ and $l_m$ in Fig. 3a), albeit for the limited data that exist, we see impressive agreement, suggesting that the structural coherence length $l_s$ corresponds closely with the characteristic electronic inhomogeneity length scale $l_e$.

An interesting question is whether the behaviour $l_e \to \xi_s(0)$, as $T \to T_c$ is evident also for *n*-type materials such as $Nd_{2-x}Ce_xCuO_4$ (NCCO)[14], even though $\xi_s(0)$ is approximately an order of magnitude larger than in *p*-type cuprate[29]. In Figure 3b) we compare the normalised phonon escape length $l_e/\xi_s(0)$ vs. $T/T_c$ for optimally doped LSCO and NCCO (both $x = 0.15$). Remarkably, the normalized $l_e/\xi_s(0)$ is very similar for NCCO and LSCO over a substantial temperature range near, and above $T_c$.

In contrast to the behaviour above and near $T_c$, the $T$-dependence of $l_e$ *well below* $T_c$ does not appear to be universal. For example $\tau_r$ (and hence also $l_e$) has previously been suggested to follow power law behaviour of the form $\tau_r \sim T^{-\eta}$ with $\nu=3$ below $T_c$ in LSCO[15], but this is apparently not always the case, as we can see from Figure 3. Expressing the escape length as $l_e/\xi_s(0) = (T/T_c)^{-\eta}$, in Fig. 3b) we observe very different behaviour for LSCO for NCCO, with $\eta \approx 2$ and $\nu \approx 0$ respectively. Furthermore, examination of data for $\tau_r$ on $YBa_2Cu_3O_{7-\delta}$[12], $Bi_2Sr_2CaCu_2O_8$[12] and $HgBa_2Ca_2Cu_3O_8$[16] also reveals quite diverse behaviour below $T_c$, which might be an indication that the typical $T$-dependence of $l_e$ might not be intrinsic, but is determined by characteristic lattice defect textures or impurities. In fact, the $T$-dependence of $\tau_r$ in high quality underdoped single crystals of $YBa_2Cu_3O_{6.5}$[24] show markedly different T-dependence than in quenched thin film samples (presumably with quenched O disorder)[11,12], the former showing much longer relaxation times (and whence longer $l_e$) below $T_c$, highlighting the importance of disorder, domains and defects, which are all likely to have a profound influence on charge inhomogeneity. Indeed, well below $T_c$, as the relaxation time $\tau_r$ increases, competing relaxation processes may become operative. Unfortunately without comparisons with other experimental data of inhomogeneity length scales below $T_c$, we cannot empirically justify the steady-state assumption in Eqs. (1) and (2) as we did for $T \geq T_c$, and we presently refrain from further analysis of $\tau_r$ only in terms of Eq.(3) well below $T_c$ until further data become available.

Although we have concentrated here on LSCO, from published data for $\tau_r$ and $\sigma(T)$ for YBa$_2$Cu$_3$O$_{7-\delta}$[12], Bi$_2$Sr$_2$CaCu$_2$O$_8$[12] and HgBa$_2$Ca$_2$Cu$_3$O$_8$[16] and comparison of $l_e$, $l_m$ and $l_s$ we can deduce that qualitatively, the behaviour is common, particularly (i) $l_e(T) \approx \xi_s(0)$ above $T_c$ (but below $T^*$) and (ii) the divergence of $l_e(T)$ around $T_c$.

In the interest of brevity, in this paper we have oversimplified on a number of issues. For instance, we have not discussed the anisotropy of $l_e$. Moreover, the use of the simple Drude model is justified mainly by the fact that $l_m$ and $l_e$ agree so well, and we have not discussed the underlying origins of its $T$-dependence. This has been discussed elsewhere[25], as was the effect of superconducting gap anisotropy on the relaxation time $\tau_r$[11,13]. In spite of this, the present analysis makes it clear that the physical properties of cuprates may be expected to be dominated by inhomogeneity, with different length scales at different $T$ and $x$.

To conclude, we suggest a physical picture which can lead to the observed behaviour of $l_e$. Doped holes are unbound at high temperature, but at some temperature $T^*$ - associated with the energy scale $kT^* \sim 2\Delta$, where $\Delta$ is the binding energy per particle - they start to form bound bipolaron pairs, leading to a charge-inhomogeneous state. Bipolarons form and dissociate according to thermal fluctuations, leading to a state which is dynamically inhomogeneous. The dimensions of these objects is determined by the balance of Coulomb repulsion and lattice attraction, as discussed in ref.[26] for example, and is of the order of the coherence length, $l$ = 1-2 nm above $T_c$. As the temperature is reduced and the density of pairs increases, these pairs start to coalesce into larger objects, which is reflected by the longer length scale $l_e$ at low temperatures[27]. When some critical temperature $T_c$, the characteristic length-scale becomes comparable to the superconducting coherence length $l(T) \approx \xi_{GL}(T)$, the percolation threshold is reached, and a macroscopically phase-coherent state is established [20].

**Acknowledgments**. The author wishes to thank V.V.Kabanov and J. Demsar for valuable discussions and suggestions, P.Kusar and J.Demsar for supplying the raw data on LSCO and S.Sugai for kindly supplying the single crystals on which the analysis is based.

**Figure 1. The photoinduced reflectivity $\Delta R/R$ measured on $La_{1.9}Sr_{0.1}CuO_4$ as a function of time for different *T*. The relaxation is described reasonably well by a single exponential. Insert: a schematic representation of the phonon escape process arising from the pairing of two particles with $E_1, k_1$ and $E_2, k_2$ (solid lines) with the emission of acoustic phonons (wavy line) in an inhomogeneous medium. The overall rate of recombination is dominated by phonon escape out of the charged volume *V*.**

**Figure 2. The phonon escape length $l_e$ as a function of temperature for $La_{2-x}Sr_xCuO_4$ for *x*=0.06, 0.1, 0.115, 0.15 and 0.2 respectively. The $T_c$ are 0, 31, 32, 38,5 and 30 K respectively. (*x*=0.06 is non-superconducting.) Here we have used $v_s$ = 4.8 nm ps$^{-1}$ for LSCO for all *x*[19].**

**Figure 3 a). A plot of $l_e$ (filled color symbols), $l_m$ (open symbols) and $l_s$ (black symbols) as a function of *T*/$T_c$ for LSCO at different doping levels. In all cases the length scale is normalized to $\xi_s(0)$[28]. The insert shows the critical behaviour just above $T_c$ for x=0.1. The mean-field prediction is plotted $l/\xi_{GL}(T) \sim (T-T_c)^{-\nu}$, with**

**ν=1/2 as a dashed line. b) The normalized escape lengths $l_e/\xi_s(0)$ for La$_{1.85}$Sr$_{0.15}$CuO$_4$ (filled squares) and Nd$_{1.85}$Ce$_{0.15}$CuO$_4$ (open circles) as a function of $T/T_c$. Here $\xi_s(0)$ = 1.8 and 18 nm respectively for LSCO and NCCO[29]. $v_s$ = 5.5 nm ps$^{-1}$ for NCCO[30].**

[1] See for example: special issues 1-2 of the *Journal of Superconductivity*: Proceedings of the International Conference on Dynamic Inhomogeneities in Complex Oxides vol. 17, pp. 1-324. (2004)**,** Guest eds. *K. A. Müller, D. Mihailovic, A. Bussmann-Holder*; D. Mihailovic and K.A.Muller, *High-Tc Superconductivity 1996: Ten years after the discovery*, **NATO ASI series** (Kluwer, 1997), Vol. **343**, p.243; A.R.Bishop et al: J.Phys.C Cond.Matt **15** L169 (2003)

[2] See for example : A.Bianconi et al Phys.Rev.Lett 76 3412 (1996), M.Acosta-Alejandro et al J.Superconductivity, **15**, 355 (2002) and references therein.

[3] Bozin et al Physica C **341**: 1793 (2000), Bozin et al., Phys.Rev.Lett. **84**, 5856 (2000)

[4] T. Egami and S.Billinge, *Underneath the Bragg Peaks. Structural Analysis of Complex Materials* (Pergamon, 2003)

[5] R.J.McQueeney et al, Phys. Rev. Lett, **82**, 628 (1999)

[6] Z.Islam et al., Phys. Rev. B **66**, 092501, (2002)

[7] H.Kimura et al, J.Phys. Soc. Jpn., **69**, 851 (2000)

[8] D.J.Derro et al., Phys Rev.Lett. **88**, 97002 (2002), S.H.Pan et al., Nature **413**, 282 (2001), C.Howald et al., Phys.Rev.B **67**, 014533 (2003), McElroy et al., Nature **422**, 592 (2003)

[9] A. Rothwarf and B.N.Taylor, Phys.Rev.Lett. **19**, 27 (1967)

[10] G.A.Sai-Halasz et al., Phys.Rev.Lett. **33**, 215 (1974), J.L.Levine and S.Y.Hsieh, Phys. Rev. Lett. **20**, 994 (1968), S.B.Kaplan et al., Phys. Rev.B **14**, 4854 (1976)

[11] V.V.Kabanov et al., Phys.Rev.B **59**, 1497 (1999)

[12] S.G.Han et al., Phys.Rev.Lett. **65**, 2708 (1990), C.J.Stevens et al.,Phys.Rev.Lett. **78**, 2212 (1997), G.L.Easley et al., Phys.Rev.Lett. **65**, 3445 (1990), D.C.Smith et al, Physica C **341**, 2219 (2000), ibid., **341,** 2221 (2000), P.Gay et al, J.Low Temp.Phys. **117**, 1025 (1999)

[13] J.Demsar et al., Phys.Rev.Lett ,**83**, 800, (1999)

[14] Y.Liu et al., Appl Phys.Lett. **63**, 979 (1993)

[15] S.Rast et al., Phys.Rev.B. **64**, 214505 (2001)

[16] J.Demsar et al, Phys.Rev.B. **63**, 54519 (2001)

[17] G.Bianchi, C.Chen and J.F.Ryan, J.Low Temp. Phys. **131**, 755 (2003)

[18] P.Kusar, J.Demsar, V.V.Kabanov, S.Sugai and D.Mihailovic, to be published.

[19] He et al, Physica B **165**, 1283 (1990)

[20] D.Mihailovic, V.V.Kabanov and K.A.Müller, Europhys. Lett. **57**, 254 (2002).

[21] V.G.Grebenik et al., Hyperfine interactions **61**, 1093 (1990)

[22] X.J.Zhou et al., Nature **423**, 398 (2003)

[23] H.Takagi et al., Phys. Rev.Lett. **69**, 2975 (1992)

[24] G.P.Segre et al., Phys. Rev.Lett. **88**, 137001 (2002)

[25] D.Mihailovic and V.V.Kabanov, J.Superconductivity **17**, 21 (2004)

[26] D.Mihailovic and V.V.Kabanov, Phys. Rev. B **63**, 054505 (2001), V.V.Kabanov and D.Mihailovic, J.Superc. **13**, 959 (2000), V.V.Kabanov and D.Mihailovic, Phys.Rev.B **65**, 212508, (2002)

[27] V.V.Kabanov, T.Mertelj and D.Mihailovic, to be published.

[28] M.Hase et al., Physica C **185-189**, 1855 (1991)

[29] G.I. Harus et al., cond-mat/9912004

[30] I.G. Kolobov et al. Phys. Rev. B **49**, 744 (1994)

# Fig.1

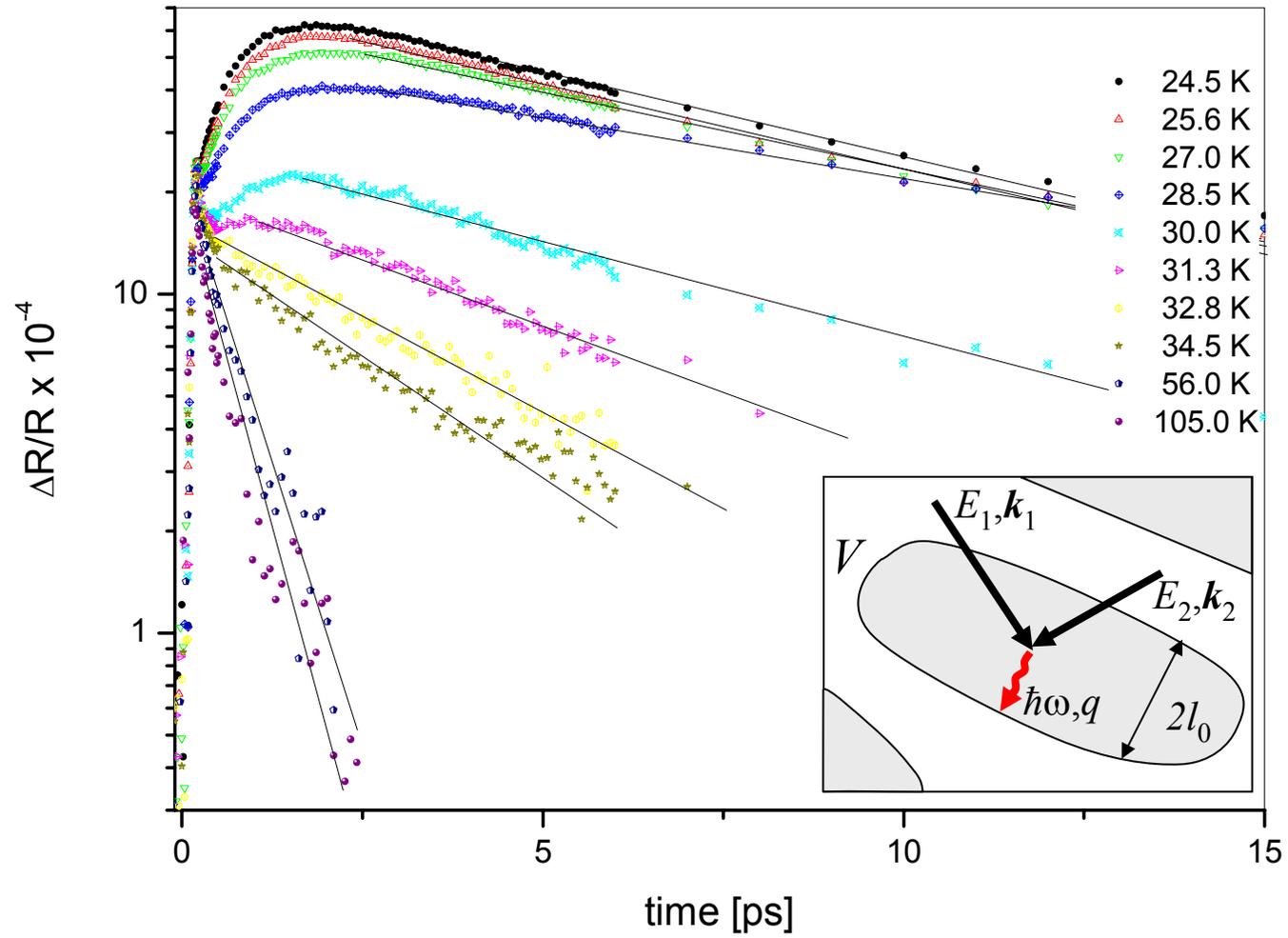

# Fig. 2

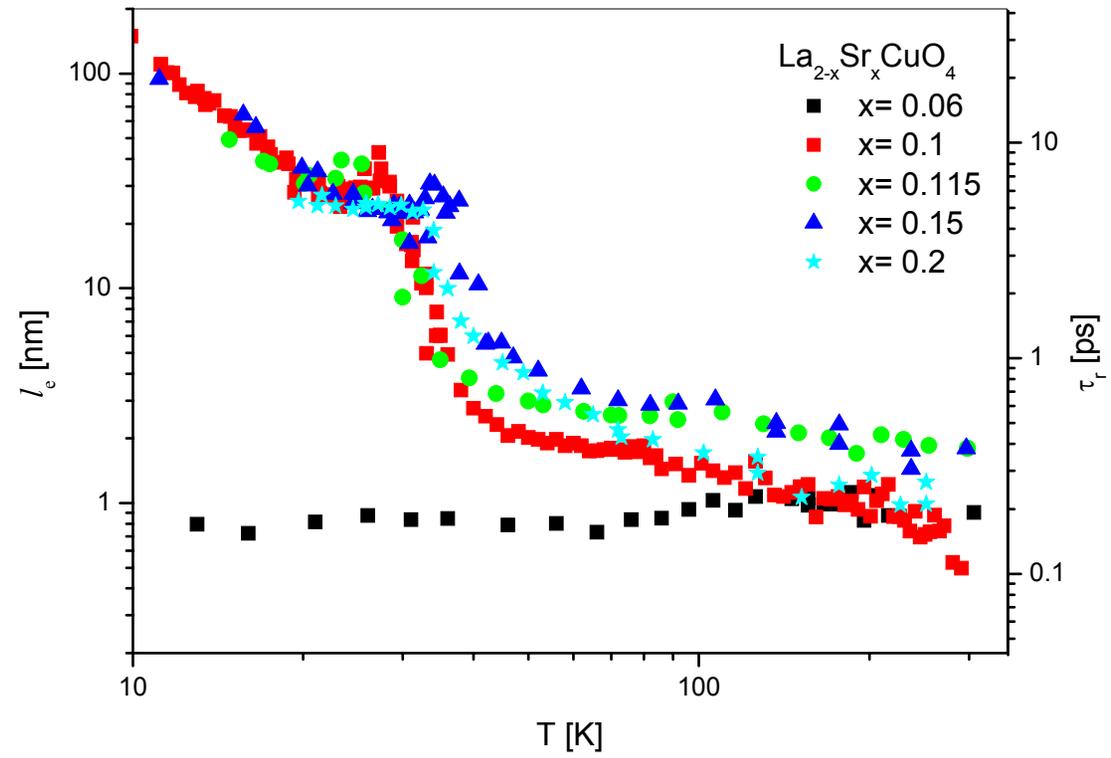

Fig. 3a)

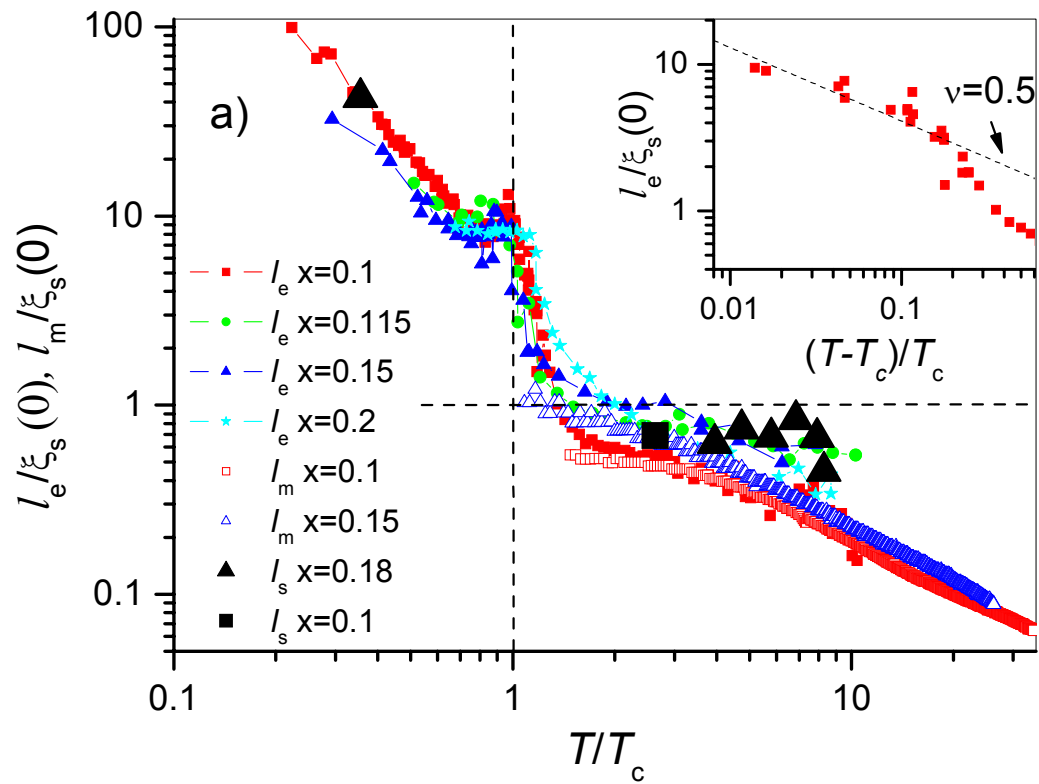

Fig. 3b)

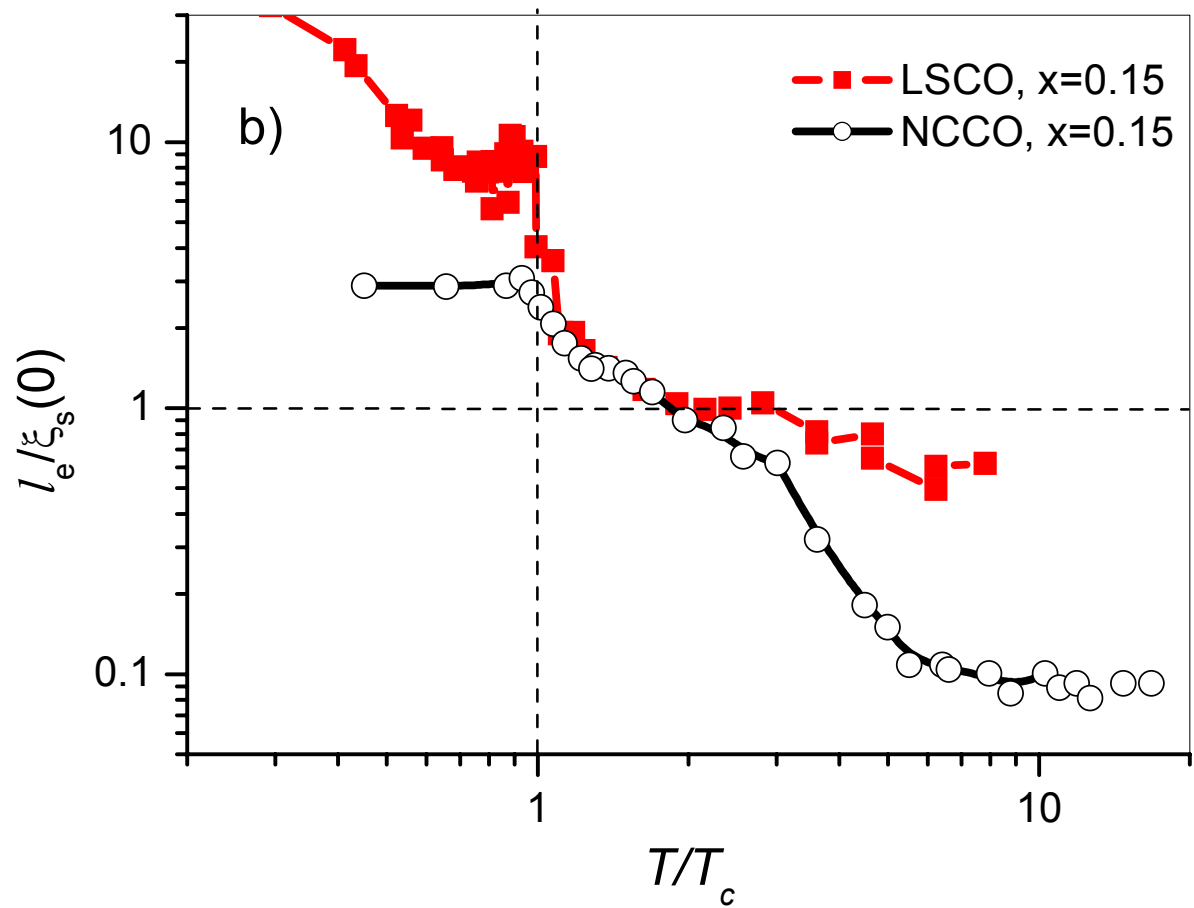